\newcommand{\refe}[1]{(\ref{#1})}
\newcommand{\citeR}[1]{Ref. [\onlinecite{#1}]}
\newcommand{\refE}[1]{Eq.~(\ref{#1})}
\newcommand{\refF}[1]{Fig.~\ref{#1}}
\newcommand{\qav}[1]{\langle #1 \rangle }
\newcommand{\rem}[1]{}
\newcommand{\ud}{\mathrm{d}}
\newcommand{\Gc}{\cal G}

\newcommand{\nablav}{\mbox{\boldmath$\nabla$}}

\newcommand{\beq}{\begin{equation}}
\newcommand{\eeq}{\end{equation}}

\newcommand{\beqa}{\begin{eqnarray}}
\newcommand{\eeqa}{\end{eqnarray}}

\documentclass[aps,prb,twocolumn,showpacs,amsmath,amssymb,floatfix]{revtex4}

\usepackage{psfig}
\usepackage{graphicx}

\begin{document}

\title{Energy dependence of current noise in double-barrier
normal-superconducting structures}

\author{G. Bignon}
\author{F. Pistolesi}
\affiliation{
Laboratoire de Physique et Mod\'elisation des Milieux Condens\'es\\
Centre Nationale de la Recherche Scientifique and
Universit\'e Joseph Fourier\\
B.P. 166 38042 Grenoble, France}
\author{M. Houzet}
\affiliation{Commissariat \`a l'\'Energie Atomique, DSM,
  D\'epartement de Recherche Fondamentale sur la Mati\`ere Condens\'ee
  SPSMS, F-38054 Grenoble, France}

\date{\today}

\begin{abstract}
We study theoretically the current-noise energy dependence for a
N-N'-S structure, where N and S stand for bulk normal metal and
superconductor, respectively, and N' for a short diffusive normal
metal.
Using quasiclassical theory of current fluctuations we obtain explicit
expressions for the noise valid for arbitrary distributions of channel
transparencies on both junctions.
The differential Fano factor turns out to depend on both junction
transparencies and the ratio of the two conductances.
We conclude that measurement of differential conductance and noise
can be used to probe the channel distribution of the interfaces.
\end{abstract}


\maketitle

\section*{Introduction}

Current noise in hybrid mesoscopic systems has been deeply
investigated in the last decade, both from the experimental and
theoretical side.\cite{blanter:2000,nazarov:2002a}
It is quite clear now that noise contains piece of information on
the charge transfer mechanism that is not present in the average
current.
The most striking example is clearly the carriers' elementary
charge, that can be obtained by measuring the noise-to-current ratio 
(Fano factor) in tunnel junctions.
As a matter of fact, in mesoscopic Normal metal/Superconducting
(N/S) hybrid structures, for energy (voltage bias and temperature)
below the superconducting gap, the elementary process responsible
for transport is Andreev reflection.\cite{blonder:1982,beenakker:1997}
It involves the transfer of {\em two} electrons at (nearly) the same
time from the superconductor to the normal metal.
This implies a {\em doubling} of the noise that has been
predicted\cite{khlus:1987,jong:1994} and 
observed.\cite{jehl:2000,kozhevnikov:2000}
The situation is particularly clear in the tunneling limit,
where the Fano-factor dependence on voltage and noise
is exactly that for a normal metal with the replacement
$e\rightarrow 2e$.\cite{pistolesi:2004}
This behavior has been recently observed in 
semiconductor/Superconductor tunnel junctions.\cite{Lefloch:2003}

N/S structures are also interesting for another reason.
If the mesoscopic structure is shorter than the coherence length,
transport is coherent and interference plays a crucial role.
Since Andreev reflection involves scattering of an electron and a
hole that are nearly time reversed particles, the random phases
acquired during the diffusion in the metal are canceled out, and
interference between electronic waves is controlled only by the
length of the path and the energy of the particles.\cite{vanwees:1992}
This leads to a strong energy (temperature or voltage bias)
dependence of the conductance that has been
predicted\cite{volkov:1993,hekking:1993,hekking:1994} and
measured.\cite{kastalsky:1991,charlat:1996}
For large energies, phases acquire a fast dependence on position and transport becomes incoherent.

Very recently, the noise was also shown to have a non-trivial
dependence on the energy. %
This dependence is {\em different} from that of the
conductance.\cite{belzig:2001,nazarov:2002b}
The cases of a long diffusive wire,\cite{belzig:2001,houzet:2004}
tunnel junction,\cite{heikkila:2002,pistolesi:2004} and double tunnel
barriers\cite{samuelsson:2003} have been considered in the literature.

The last structure is particularly interesting since interference is
enhanced by increasing the number of reflections.
A Fabry-Perrot structure made of two barriers between the
superconductor and the normal metal is expected to show a strong
energy dependence conductance.
This was predicted some time ago\cite{volkov:1993} for N-I-N'-I-S
structures (where I is an insulating barrier) using quasiclassical
Green's function approach, and then confirmed
experimentally.\cite{quirion:2002}
More recently the noise in this tunneling structure has been
calculated.\cite{samuelsson:2003}
The tunneling condition greatly simplifies the theoretical approach.
This assumption does not limit severely the range of the normal-state
conductances that can be theoretically investigated since the number
of channels in most cases is very large.
However, for given normal-state conductances 
one expects a dependence of current and noise on the 
actual value of the transparencies. 
Concerning the current, this was confirmed by the work of Clerk {\em
et al.}\cite{clerk:2000} where the conductance for non-tunnel N-N'-S
structures has been evaluated by means of random matrix theory.
The behavior of the noise when the interfaces are not tunneling is
the object of the present work.

In this paper we calculate the current noise for a N-N'-S structure
without restrictions on the distribution of channel transparencies on
both interfaces.
We use quasiclassical Green's function
technique\cite{eilenberger,larkin:1968,usadel:1970}
with boundary conditions modified by the introduction of a 
counting field\cite{levitov:1996,nazarov:1999b,BelzigNazarovFCS:2001}
allowing to calculate the noise.
Exploiting the parametrization for the Green's function proposed by
two of the authors in Ref. \onlinecite{houzet:2004} we obtain the
expressions for the voltage and temperature dependence of current
noise in terms of a complex parameter to be found numerically.
In some limiting cases the calculation can be performed to the end
analytically.
In all others the numerics is straightforward.
We find that when the conductances are of the same order of magnitude,
the channel distribution becomes crucial for the determination of the
energy dependence of both the current and noise.
The expressions we provide can be used to characterize interfaces
when current and noise can be measured accurately.
Even if this is non trivial from the experimental point of view, one
should consider that it is very difficult to control only by means 
of the fabrication the transparency of an interface, {\em i.e.} the value
of the transparencies and their distribution.
If the average transparency can be easily estimated from the size of
the contact, the true distribution remains out of the reach of any
probe.
That is why having a theory that predicts the conductance and noise
for an arbitrary distribution of the channel transparencies can be a
useful tool.

The paper is organized as follows.
In Sec.~\ref{section:model} we introduce the circuit theory model
and derive the main equations. 
In Sec.~\ref{section:noise} we obtain the expressions for the 
current and the noise. We discuss our results for different 
distribution of channel transparencies.
Sec.~\ref{conclusions} gives our conclusions.

\section{Model and basic equations}
\label{section:model}

\begin{figure}[tbh]
\centerline{
\includegraphics{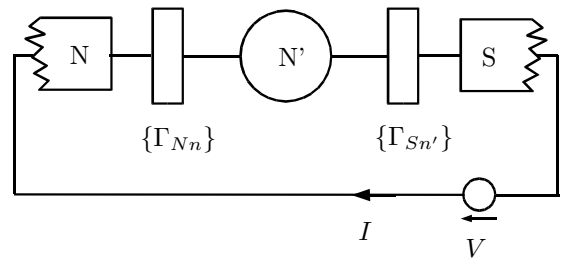}}
\caption{\label{fig1} Schematic picture of
the N-N'-S junction. $\{\Gamma_{Nn }\}$ and $\{\Gamma_{Sn'}\}$ are
channel transparencies of the N-N' and N'-S barriers.}
\end{figure}

We consider a N-N'-S structures with two junctions characterized by
their set of channel transparencies: $ \{ \Gamma_{N n} \} $ for the
N-N' barrier and $\{\Gamma_{S n'} \}$ for the N'-S barrier, $n$ and
$n'$ being channel labels (see Fig.~\ref{fig1}).
Consequently, the conductances are $g_{N(S)} = g_Q \sum_n
\Gamma_{N(S)n}$, where $g_Q =2 e^2/h$ is the quantum of conductance.
We assume that $g_{N/S}$ is large enough to completely neglect
the voltage drop in the N' part.
Namely, we require that the time necessary for an electron incoming
from the leads to visit the whole N' region (dwelling time $\tau_D$) is
much smaller than the time spent in the region itself (escape
time $\tau$).
This corresponds to asking that the Thouless energy
$E_{\textrm{Th}} 
\equiv  
\hbar/\tau_D 
= 
\hbar D/L^2$ ($D$ being the diffusive constant
and $L$ the typical size of N') is much larger than
$E_{\tau} \equiv \hbar/\tau = (g_N+g_S)\delta/(4 \pi  \, g_Q)$ 
($\delta$ being the average level spacing for N').
We also assume that $L\ll\xi_d=\sqrt{\hbar D/\Delta}$ (or
equivalently $E_{\textrm{Th}}\gg \Delta$), where $\Delta$ is the
superconducting gap of S, so that the spatial dependence of the
proximity effect can be neglected in N'.

Proximity effect is thus completely controlled by $E_{\tau}$ and
charge transport does not depend on the shape of N'.
Hence we can consider N' as an isotropic zero dimensional conductor.
We also assume that $g_{N/S} \gg g_Q$: Each barrier has a large number
of conduction channels.
Coulomb blockade and weak localization effects are then negligible.
Finally we require that the escape time is much smaller than phase
breaking and inelastic time.
All these requirements are met, for instance, in the experiment of
Ref. \onlinecite{quirion:2002}.

Within these assumptions, one can apply the so-called ``circuit
theory'' to calculate current, noise and higher current
cumulants.\cite{nazarov:1999a,nazarov:1999b,BelzigNazarovFCS:2001,belzig:2002,nazarov:2002b}
In particular the central region can be approximated with a single
node, since any internal spatial dependence is negligible.
The conductor is thus discretized into three nodes connected via two
connectors, see Fig.~\ref{fig1}.
Each node is characterized by a quasiclassical matrix Green's function
in the Keldysh($\hat{\,}$)-Nambu($\bar{\,}$) space, $\check{G}_{N/S}$
for N and S leads and $\check{G}$ for N' depending on the energy $E$
and a counting field~$\chi$.\cite{levitov:1996}

The counting field appears as a modification of the
boundary conditions.
In our case this corresponds to transforming the normal reservoir
Green's function as follows:\cite{BelzigNazarovFCS:2001}
\beq
  \check{G}_{N} (\chi)
  =
  e^{i \chi \check{\tau}_K/ 2} \ \check{G}_{N0}
  \ e^{ - i \chi \check{\tau}_K / 2},
\eeq
where $\bar\sigma_i,\hat\tau_{j(i,j=1,2,3)}$ are Pauli matrices,
$\check\tau_{K}=\hat\tau_3\otimes\bar\sigma_1$, and $\check{G}_{N0}$
is the normal metal quasiclassical Green's function in the diffusive
limit (for a recent review see \citeR{kopnin:2001}):
\beq
    \check{G}_{N0}
    =
    \left(
    \begin{array}{cc}
    \hat{\tau}_3 & 2(f_{T0}+f_{L0} \hat{\tau}_3 ) \\
    0 &    - \hat{\tau}_3
    \end{array}
    \right).
\eeq
Here, $ f_{T0} = f(E - eV) -f(E + eV)$, $ f_{L0} = 1-f(E + eV) -f(E
-eV)$, $f$ is the Fermi function at temperature $T$, and $V$ is the
voltage bias between the normal metal and the superconducting
reservoir.

The Green's function in the superconducting reservoir is
\beq
\check{G}_S = \left(
\begin{array}{cc}
   \hat{R}_{S} & \hat{K}_{S} \\
   0 &   \hat{A}_{S}
\end{array}
\right).
\eeq
Here, $\hat{R}_{S} $ is the retarded part given by:
\beq
 \hat{R}_{S} = \frac{1}{\sqrt{(E+i \eta)^2 -|\Delta}|^2} \left(
\begin{array}{cc}
   E & \Delta \\
   - \Delta^{\star} &  - E
\end{array}
\right)
\eeq
with the branch cut of the square root going from $-\Delta$ to $+\Delta$ on
the real $E$ axis.
The advanced part is given by $\hat{A}_{S}= -\hat{\sigma}_3
\hat{R}^{\dag}_{S} \hat{\sigma}_3$, and
the Keldysh part follows by the equilibrium condition of the
reservoir: $\hat{K}_{S} = (f(E)-f(-E)) (\hat{A}_{S}-\hat{R}_{S})$.
In the following, we focus on the supgap regime, so we can limit
ourselves to $E \ll |\Delta|$.
Moreover, since there is only one superconductor in the problem, we
can choose $\Delta$ real.
Then the matrix Green's function of the superconductor simplifies to
$\check{G}_{S} = \hat{\tau}_2 \otimes \bar{1}$.

The Green's function in the central node satisfies the normalization
condition
$\check{G}^2 = \check{1}$
and the symmetry property:\cite{houzet:2004,footnotesymmetry}
\beq
 \label{equation:symmetry}
 \check{G}^{\dag}(-\chi)
 =
 - \check{\tau}_L \ \check{G}(\chi) \  \check{\tau}_L
\eeq
with
$\check{\tau}_L = \hat{\tau}_3 \otimes \bar{\sigma}_2 $.
(Similar relations hold for $\check{G}_{N/S}$ as well.)
It is solution of the Usadel equation:\cite{usadel:1970,kopnin:2001}
\beqa
\label{equation:usadel}
\hbar D \mathbf{\nablav} \left( \check{G} \mathbf{\nablav}  \check{G} \right)
 -  i E [ \check{G}_E , \check{G} ] = 0  \, ,
\qquad \check{G}_E = \hat{\tau}_3\otimes \bar{1} \, .
\eeqa
We integrate this equation over the volume $\mathcal{V}$ of N'. Using the divergence 
theorem, it gives:
\beqa
\int_{\partial \mathcal{V}} d^2\mathbf{S} \cdot \left(\sigma_0 \check{G} \mathbf{\nablav}
  \check{G} \right) -2 \, i \frac{e^2 \nu_0 \mathcal{V} \, E}{\hbar} \, 
  [ \check{G}_E , \check{G} ] = 0 \, ,
\eeqa
where $\nu_0$ is the density of states per spin of N', and 
$\sigma_0= 2 e^2 D \nu_0$  its conductivity  in the normal metallic state. 
Using boundary conditions for the Green's functions over 
the surface $\partial V$ of the grain,\cite{zaitsev:1984,nazarov:1999a} we have:
\beqa
- \int_{\partial \mathcal{V}} d^2\mathbf{S} . \left(\sigma_0 \check{G} \mathbf{\nablav}
  \check{G} \right) = \check{I}_N + \check{I}_S  
\eeqa
with 
\begin{subequations}
\label{equation:allmatrixcurrent}
\beqa
    \check{I}_N & = & g_Q  \sum_{n} \frac{2 \, \Gamma_{Nn}  [
    \check{G}_{N}(\chi) , \check{G} (\chi)]}{ 4 +  \Gamma_{Nn}  ( \{
    \check{G}_{N}(\chi) , \check{G}(\chi)  \} - 2 )  },
    \label{equation:allmatrixcurrent_1}
    \\
    \check{I}_S & = & g_Q  \sum_{n} \frac{2 \, \Gamma_{Sn}  [
    \check{G}_{S} , \check{G} (\chi)]}{ 4 +  \Gamma_{Sn}  ( \{
    \check{G}_{S} , \check{G}(\chi)  \} - 2 )  } \, .
    \label{equation:allmatrixcurrent_2}
\eeqa
\end{subequations}
Then $\check{G}$ is fully determined by \refE{equation:usadel} which takes
the form of a conservation-like equation for the spectral matrix current:
 \beq
 \label{equation:matrixcurrent}
 \check{I}_N + \check{I}_S
    +\check{I}_E = 0 \,  ,
 \eeq
where
\beqa
    \check{I}_E & = & g_Q \frac{ 2 i \pi  \, E }{\delta}
    [  \check{G}_{E} , \check{G}(\chi) ] \, .
\eeqa
Here $\check{I}_E$ is the
``leakage'' matrix current,\cite{nazarov:1999a} which takes into
account the relative dephasing between electron and hole during
their propagation in the central node N', whose mean level spacing
is $\delta= 1/ ( \nu_0 V) $.
The estimate for the inverse escape time, $E_{\tau}/\hbar = 
\delta (g_N +g_S)/(4 \pi \hbar \, g_Q )$, follows from comparison between the
amplitudes of $\check{I}_N+\check{I}_S$ and $\check{I}_E$.

Once the matrix $\check{G}(\chi)$ is known, current, zero frequency
noise and all higher current cumulants can be obtained by
differentiation of $I(\chi)$ defined as follows:
\beq
    \label{Ichi}
    I(\chi) 
    = 
    - \frac{1}{8e} \int \ud E \ \textrm{tr}  [
    \check{\tau}_K  \check{I}_N ] 
    \, .
\eeq
(By matrix current conservation \refe{equation:matrixcurrent} $I(\chi)$
 equals minus expression \refe{Ichi} with $\check{I}_N$ substituted by
  $\check{I}_S$.)
The first two moments are the average current,
\beq
   \label{courant_general}
   I =  I(\chi) \big|_{\chi=0} \,,
\eeq
and the current noise,
\beq
   \label{bruit_general}
   S = 2 i e
   \left.\frac{ \partial I ( \chi ) }{\partial \chi}
   \right|_{\chi=0} \, .
\eeq

For tunneling interfaces, $\Gamma_n\ll1$, the boundary conditions
simplifies since one can neglect the anticommutator in the
denominator of Eqs.~(\ref{equation:allmatrixcurrent_1}) and
(\ref{equation:allmatrixcurrent_2}).
In that limit the matrix $\check{G}(\chi)$ can be found
analytically.\cite{borlin:2002,samuelsson:2003}
It is thus possible to study not only the
current and noise, but the whole set of cumulants.
In the general case of arbitrary value of $\Gamma_n$ there is no
analytical solution available for $\check{G}(\chi)$.

If one restricts to the first two cumulants, $I$ and $S$, which are
more accessible experimentally, it is possible to write simplified
equations for the coefficient of the expansion of
$\check{G}$ in $\chi$:\cite{belzig:2001,houzet:2004}
\beq
    \label{chiExp}
    \check{G}(\chi) = \check{G}_0 - i \frac{\chi}{2} \check{G}_1 +
    \mathcal{O} (\chi^2)
    \,.
\eeq
Finding $\check{G}_0$ gives the current while
$\check{G}_1$ leads to the noise.
In the following we follow this program and solve
\refe{equation:matrixcurrent} for the first two
orders in $\chi$.

\section{Current and Noise}
\label{section:noise}

\subsection{Current evaluation}

To obtain the current one has to evaluate \refE{Ichi}.
For this, we need $\check{G}_0$ as defined in \refE{chiExp}.
A crucial step to solve the problem is to take into account the
normalization condition without redundancy in the parametrization.
When the counting field vanishes, the solution is well known and it
consists in the following parametrization of $\check{G}_0$:
\beq
   \check{G}_0 = \left(
   \begin{array}{cc}
   \hat{R} & \hat{K} \\
   0 &   \hat{A}
   \end{array}
   \right)
\eeq
with
\begin{subequations}
\beqa
  \hat{R} 
  & = &  
  \hat{\tau}_3 \cosh \theta + i  \hat{\tau}_2
  \sinh \theta  \,, 
  \quad \hat{A}  
  =  
  - \hat{\tau}_3 \ \hat{R}^{\dag} \   \hat{\tau}_3 \, , 
  \\
  \hat{K} 
  & =  & 
  \hat{R}  \hat{f} - \hat{f} \hat{A},  \qquad \hat{f} 
  = f_L + \hat{\tau}_3 f_T
  \,
  .
\eeqa
\end{subequations}
Here, the parameters $f_T$ and $f_L$ are real, as follows from
Eq.~(\ref{equation:symmetry}) at $\chi=0$. The complex number 
$\theta$ characterizes the paring in the grain: $\theta = -i \pi /2$
corresponds to a fully superconducting state and $\theta=0$ to a normal one.
Substituting this form for $\check G_0$ into
\refE{equation:matrixcurrent} at $\chi = 0$ one can determine
$\theta$, $f_L$, and $f_T$.
The retarded or advanced parts give the equation for $\theta$:
\beqa
  \label{equation:theta}
  g_N  \qav{Z_N }  \sinh \theta  + i \varepsilon g_D  \sinh \theta +
  i  g_S \qav{Z_S}  \cosh \theta= 0,
\eeqa
where $\varepsilon = E / E_{\tau}$, $g_D = g_N +g_S$, 
$
Z_N=[1+\Gamma_N ( \cosh \theta -1 )/2]^{-1}
$ 
and 
$ Z_S = [1+ \Gamma_S ( i \sinh \theta -1 )/2]^{-1} $.
Here,
\beq
  \qav{f(\Gamma_{\alpha})}=
   {\sum_n
   \Gamma_{\alpha n} f(\Gamma_{\alpha n})
   \over
   \sum_n \Gamma_{\alpha n}
   } \,
\eeq
stands for the average over channel transparencies with $\alpha =$ N or S.
The Keldysh part of the spectral-current-conservation equation
(\ref{equation:matrixcurrent}) gives $f_L = f_{L0} $ and $f_T /
f_{T0} = c $ with
\beqa
  c = \frac{ g_N \tanh \theta_1}{2\, \varepsilon g_D \sin \theta_2}
  \qav{ \left[ 
  (2-\Gamma_N ) \, \cos \theta_2 + \Gamma_N \, \cosh
  \theta_1 \right] | Z_N |^2}.
\label{cdef}
\nonumber \\
\eeqa
Here, we used the decomposition $\theta=\theta_1+i\theta_2$ into
real and imaginary part. Finally, the mean current is given by
 \beq
    I =
    \frac{1}{ 2 e}
    \int^{\infty}_{-\infty} \ud E \, f_{T0} \,  \mathcal{G} (E)
\eeq
with
\beqa
\label{equation:current}
    \mathcal{G} (E)  & = &
    c \, g_S  \, \cosh \theta_1 \nonumber \\
    && \times  \,
    \qav{
    \large[
    -\sin \theta_2 + \Gamma_S \, (
    \cosh \theta_1 \, + \, \sin \theta_2
    )/2 \large]) | Z_S  |^2} \,.
 \nonumber   \\
\eeqa
At zero temperature the differential conductance $G \equiv \ud I/\ud
V$ equals $\mathcal{G} (eV)$.
For uniform transparency, expression \refe{equation:current}
coincides with that obtained by \citeauthor{clerk:2000} in
Ref.~\onlinecite{clerk:2000} using random matrix theory.

We now discuss the conductance for small and large energy.
Let us begin with the low energy limit $ eV \ll E_{\tau}$, {\em
i.e.}, the completely coherent case.
$\theta$ is then an imaginary number.
\refE{equation:current} reduces to:
\beq
    G^{-1}_{\textrm{coh}}
    =
    (\widetilde{g}_N)^{-1} + (\widetilde{g}_S)^{-1}
    \label{equation:courant_coherent} \, ,
\eeq
where
\beqa
    \widetilde{g}_N &
    =
    &
    g_N \cos \alpha \qav{ \widetilde{Z}^2_N } +
    g_N \qav{ \Gamma_N \widetilde{Z}^2_N } ( 1 -\cos \alpha) / 2 \, ,
\nonumber
\\
   \widetilde{g}_S & = & g_S \sin \alpha \qav{\widetilde{Z}^2_S }+ g_S
   \qav{ \Gamma_S \widetilde{Z}^2_S } ( 1 -\sin \alpha)/2 \,,
\nonumber
\eeqa
with $\widetilde{Z}_N = [1+ \Gamma_N ( \cos \alpha -1 )/2]^{-1}$
and $\widetilde{Z}_S = [1+ \Gamma_S ( \sin \alpha -1 )/2]^{-1}$.
The real parameter $\alpha = - \textrm{Im}(\theta)$ is the solution 
of the equation:
\beqa
\label{equation_alpha}
g_N\, \sin \alpha \,
\qav{\widetilde{Z}_N } = g_S \, \cos \alpha \, \qav{\widetilde{Z}_S}\, .
\eeqa
Coherent conductance strongly depends on the ratio $g_N/g_S$.
When the central island is well connected to N ($g_N \gg g_S$) 
$\theta=0$. The grain is in the normal state. Then differential 
conductance is given by
$G_{\textrm{coh}} = g^{\textrm{And}}_{S} = 2 g_S \qav{ \Gamma_S
/(2-\Gamma_S)^2}$:
the charge transfer is dominated by Andreev reflection at N'-S
interface.\cite{beenakker:1992}
In the opposite case of an island well connected to S ($g_N \ll g_S$) 
$\theta= -i \pi/2$, the grain is superconducting and  we have
$
G_{\textrm{coh}} = g^{\textrm{And}}_{N} = 2 g_N \qav{
\Gamma_N/(2-\Gamma_N)^2}$.
This means that conductance is dominated by Andreev reflection at N-N' barrier.
We can also note that $G_{\textrm{coh}}$ is invariant under the
transformation $ \{\Gamma_{Nn } \} \leftrightarrow \{ \Gamma_{Sn' }
\} $ in \refE{equation:courant_coherent}.
Thus when an electron crosses the N-N'-S structure, it can not
distinguish which barrier is closer to the superconductor.

\begin{figure}[tbh]
\centerline{
\includegraphics[width=7cm]{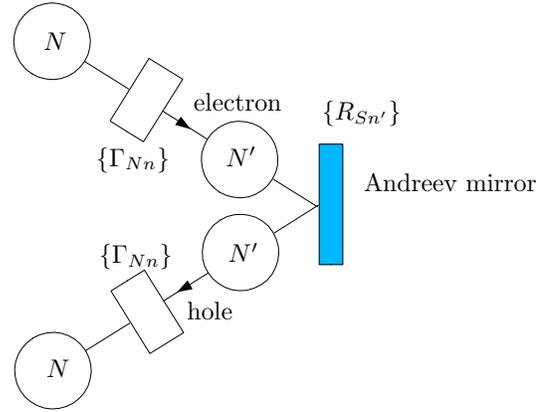}}
\caption{\label{fig:incoh}
Schematic picture of the N-N'-S junction in the incoherent
regime. Electrons are Andreev reflected into hole at N'-S barrier.}
\end{figure}

In the opposite limit of $e V \gg E_{\tau}$ transport is incoherent. The large energy mismatch between electrons and Andreev reflected holes washes out interference effects.
We find the following expression for the conductance
\beq
  \label{equation:cond_incoh}
  G_{\textrm{class}}^{-1}
  =
  g_N^{-1}  + (g^{\textrm{And}}_S )^{-1}
  \,,
\eeq
that is now no more invariant for exchange of the N-S and 
N'-S barriers. The grain is in the normal state ($\theta=0$).
The physical interpretation for the incoherent transport
is simple since one can treat one channel at the time
(electrons do not interfere).
For a Cooper pair to be transferred across the double barrier
structure the electron has to undergo the following
steps: crossing of the N-N' barrier, Andreev
conversion to a hole at the N'-S junction (with probability
$ R_{Sn'} = \Gamma^{2}_{Sn'}/(2-\Gamma_{Sn'})^2$ per channel),
and finally crossing of the N-N' barrier (see Fig.~\ref{fig:incoh}).
Thus in the incoherent limit, the double junction is equivalent to
three junctions in series of transparencies $\{ \Gamma_{Nn} \}$, $\{
R_{Sn'} \}$ and $\{ \Gamma_{Nn} \}$, respectively, with an
elementary transferred charge $2e$.
Conductance is then given by Ohm's law for the three conductances in
series multiplied by a factor two: $G_{\textrm{class}} = 2 [ (g_Q
\sum_n \Gamma_{Nn} )^{-1} + (g_Q \sum_{n'} R_{Sn'} )^{-1} +(g_Q
\sum_n \Gamma_{Nn} )^{-1} ]^{-1}$, which coincides with
\refE{equation:cond_incoh}.

\begin{figure}[tbh]
\centerline{
\includegraphics[scale=0.8]{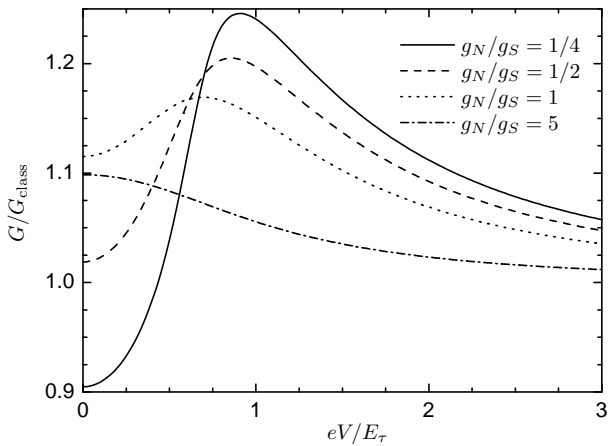}}
\caption{\label{fig:cond_dirty}
Differential conductance normalized by its classical value as a
function of $eV/E_{\tau}$ for two disordered interfaces at zero
temperature.
A peak appears near $eV \approx E_{\tau}$ when $g_N \approx g_S$ and
it becomes sharper for $g_N/g_S\rightarrow 0$.}
\end{figure}

For intermediate energies, the shape of $G$ depends both on
the ratio of the two conductances and the set of channel transparencies.
A particularly relevant case of channel distribution is that
of a disordered interface:\cite{schep:1997}
\beq
    \rho_{\alpha}(\Gamma) \equiv \sum_n \delta(\Gamma-\Gamma_{\alpha n})
    =
    \frac{g_{\alpha}}{g_Q \, \pi} \frac{1}{\Gamma^{3/2}\sqrt{1-\Gamma}}
    \,,
\eeq
where $\alpha=$N or S.
We plot the conductance for this case and different
values of the ration $g_S/g_N$ in Fig.~\ref{fig:cond_dirty}.
Qualitatively one sees a cross-over from a ``reflectionless''
tunneling behavior, typical of tunnel junctions (with a zero-bias
peak) to a ``re-entrant'' behavior with a peak at $eV$ of the order
of $E_\tau$.
In both cases the qualitative explanation is simple.
In the tunnel case, the electron tries many times to enter the
superconductor.
At low energy, the corresponding quantum paths add coherently,
giving a large resulting current.
Any increase in the energy reduces the coherent contribution to the
current since interference is suppressed and, thus, the mixed terms
vanish.
This explains the zero-bias peak in the $G(V)$ plot.
On the other hand, when the superconducting barrier is transparent
the electrons always succeed in being converted to holes, but
Andreev reflection comes with a phase factor ($-i$) that induces
destructive interference among electronic waves for
$E=0$.\cite{beenakker:2000}
The loss of coherence among waves can thus enhance the current
leading to a maximum in the $G(V)$ plot.
This behavior is very similar to the one observed in a diffusive
wire.\cite{volkov:1993}
One sees nevertheless that the effect is much larger here, since the
Fabry-Perot structure enhances interference.
We will discuss the role of barrier transparencies in \ref{barriers}.

\subsection{Noise evaluation}

Let us now consider the main subject of the paper: the noise.
As stated above we need to solve \refE{equation:matrixcurrent} in
first order in $\chi$.
We expand thus each spectral current:
$
\check{I}_{\beta} 
= 
\check{I}^{0}_{\beta} + \chi \check{I}^{1}_{\beta} + \mathcal{O} ( \chi^2 )
$ 
with
$\beta =$N, S, or E.
We obtain:
\beqa
\check{I}^{1}_{N}
& = & i g_N \big( \qav{\check{D}_{N}} ( [\check{G}_{N1} , \check{G}_0
] - [\check{G}_{N0} , \check{G}_1 ] ) \nonumber \\
&& - \qav{ \Gamma_{N} [\check{G}_{N0} , \check{G}_0 ] \check{D}_{N} (
\{ \check{G}_{N1}, \check{G}_{0} \} - \{ \check{G}_{N0}, \check{G}_{1}
\} ) \check{D}_{N}} \big)
\nonumber \, , \\
\check{I}^{1}_{S} & = &
-i g_S \big( \qav{\check{D}_{S}} [\check{G}_{S} , \check{G}_1 ]
\nonumber \\ && - \qav{\Gamma_{S} [\check{G}_S , \check{G}_0 ]
\check{D}_{S} \{
\check{G}_{S} , \check{G}_1 \} \check{D}_{S}} \big)
\nonumber \, , \\
\check{I}^{1}_{E} & = &
\frac{g_D \varepsilon}{4 } [\check{G}_E , \check{G}_1 ] \nonumber \, ,
\eeqa
with
$ \check{D}_{\alpha } = ( 4 + \Gamma_{\alpha } ( \{
\check{G}_{\alpha 0} , \check{G}_0 \} - 2 ) )^{-1 }$,
$ \alpha =$N or S,
and $\check{G}_{N1} = [ \check{\tau}_K , \check{G}_{N0} ]$.
Zero frequency current noise is given by
\beq
\label{equation:noise1}
 S = \frac{i}{4} \int \ud E \ \textrm{tr} [\check{\tau}_K
 \check{I}^1_S]
 \,.
\eeq
Here, the unknown matrix $\check{G}_1$, cf. \refE{chiExp}, satisfies
\beqa
\label{equation:current_cons1} \check{I}^1_N  + \check{I}^1_S +
\check{I}^1_E = 0 \, .
\eeqa
Additionally the normalization of $\check G$ implies $ \{ \check{G}_0
, \check{G}_1 \} =0$.
This can be satisfied by defining
 $\check{G}_1 = [ \check{G}_0 , \check{\phi}]$ for any
$\check{\phi}$.
We use the parametrization found in Ref. \onlinecite{houzet:2004}
for the matrix $\check \phi$:
\beq
\check{\phi} = \left(
\begin{array}{cc}
   a f_{T0} \hat{\tau}_1 - c \hat{f} \hat{\tau}_3 & b \hat{\tau}_3 +d \\
   c \hat{\tau}_3 &   a^\ast f_{T0} \hat{\tau}_1 + c \hat{f} \hat{\tau}_3
\end{array}
\right)  \, .
\eeq
The symmetry condition \refe{equation:symmetry} on $\check G$
implies that $\check{\phi}^{\dag} = - \check{\tau}_L \check\phi
\check{\tau}_L$; it follows that $b$, $c$, and $d$ are real, while
$a$ is complex.
The parameter $c$ has been already given in \refE{cdef}.
Inserting this form for $\check\phi$ into
\refE{equation:current_cons1} one obtains a complete set of equations
for all the parameters of $\check{\phi}$.
The equation for $a$ is given by the antidiagonal elements of the retarded part of
\refE{equation:current_cons1}.
The Keldysh part of the same equation gives the equations for $b$ and $d$.
Finally, using \refE{equation:noise1}, zero frequency current noise
takes the form:
\beq
  \label{equation:noise2}
S = \int \ud E \, \, \{ \mathcal{G}(E)
   [ 1 -
     f^{2}_{L0}] + \mathcal{S}_{T} (E) \, f^{2}_{T0} \}
  \,.
\eeq
The rather cumbersome expressions for $a$, $b$, $d$, and
$\mathcal{S}_{T}$ are given in the Appendix.
Here we only stress that the analytic expressions for the
coefficients all depend on a single complex number, $\theta$,
solution of \refE{equation:theta}.
Even if $\theta$ is given by the solution of an algebraic equation
it is not always possible to obtain an analytical 
expression for it.
Nevertheless, once this parameter is known numerically, it is enough
to substitute it into the expressions given in the Appendix to 
obtain the value of the current noise.
Note that knowledge of $\theta$ is already necessary to obtain the
conductance.

Let us now discuss the result in some details.
We first note that \refE{equation:noise2} for $eV \ll k_B T$ 
correctly agrees with the fluctuation-dissipation theorem.\cite{callen:1951} 
As a matter of fact, in this case $f_T=0$ and the remainder gives
precisely $S = 4 k_B T \, G(T)$.
In the opposite limit noise is not simply related to the conductance
and has to be computed with \refE{equation:noise2}.
In the zero temperature limit ($k_B T \ll eV$, $E_\tau$) the
experimentally accessible differential Fano factor becomes from
\refE{equation:noise2}:
\beq
  \label{equation:fano_diff}
  F(V)
  \equiv
  \frac{1}{2 e G(V)} \, \frac{ \ud S(V) }{ \ud V }
  =
  1+\frac{\mathcal{S}_T(eV)}{\mathcal{G}(eV)}
  \,.
\eeq

Let us now discuss as in the conductance case the two analytically
tractable limits: the completely coherent and incoherent cases.
In the coherent limit one can obtain closed analytical 
expression for the noise depending on the parameter $\alpha$
solution of \refE{equation_alpha}. However they are rather cumbersome and 
we will not show them. In the specific case of  two transparent barriers, 
we recover the recent analytical result of Vanevi\'{c} {\it et al.} 
 in \citeR{vanevic:2004}.
Similarly to the conductance, the expression for the Fano factor is left unchanged 
when the set of transparencies of the two barriers are exchanged:
 $\{\Gamma_{N n}\} \leftrightarrow \{\Gamma_{S n'}\} $.
The Fano factor depends on the ratio $g_N /g_S$. If the grain is well
connected to the superconductor ($g_N \ll g_S$), $F_{\textrm{coh}} = 
2 \sum_{n'} R_{Sn'} (1 -R_{Sn'} ) / \sum_{n'} R_{Sn'}$: we obtain the Fano factor
of N'-S interface alone.\cite{jong:1994} In the opposite limit, $g_N \gg g_S$,
 $F_{\textrm{coh}} = 2 \sum_n R_{Nn} (1 -R_{Nn} ) / \sum_n R_{Nn}$:
the Andreev reflection occurs at N'-N barrier. It is interesting to notice
that even if transport properties ($G_{\textrm{coh}}$ and $F_{\textrm{coh}}$)
do not depend on the relative position of the barriers, the state of the grain does.
It can be normal or fully superconducting depending on $g_N/g_S$. 

We consider now the incoherent limit: $eV\gg E_\tau$.
From \refE{equation:fano_diff} 
we find the following explicit form for the differential Fano factor
\begin{widetext}
\beqa
    F_{\textrm{class}} & = & \biggl[ (g^{\textrm{And}}_S)^3
    \frac{\sum_n \Gamma_{Nn}\,(1-\Gamma_{Nn} /2)}{{\sum_n \Gamma_{Nn}
    }} + 2 g_N g^{\textrm{And}}_S (g_N + g^{\textrm{And}}_S) + 2 g^3_N
    \frac{ \sum_n R_{Sn}\,(1-R_{Sn})}{{\sum_n R_{Sn} }} \biggr]
    \frac{1}{(g_N + g^{\textrm{And}}_S)^3 }.
\label{fano_incoh}
\eeqa
\end{widetext}
This result can also be found using the technique developed by
\citeauthor{belzig:2003epl}.\cite{belzig:2003epl}
The physical interpretation is the same described for 
$G_{\textrm{class}}$, the only difference is that here we 
need to calculate the current fluctuation at each barrier
instead of the current.
Indeed, in the classical limit, the structures can be schematized
as a series of three junctions of transparencies
$\{ \Gamma_{Nn} \}$, $\{ R_{Sn'} \}$, and
$\{\Gamma_{Nn} \}$ with decoherent cavities in between.
Again the elementary transferred charge is $2e$ (see Fig.\ref{fig:incoh}).
The Fano factor for a series of two junctions separated by a
decoherent cavity has been evaluated (for elementary charge
$e$):\cite{nazarov:2002b,nagaev:2002}
\beqa
   F_{12}(g_1,F_1;g_2,F_2)
   =
   \frac{g^{3}_1 F_2 +g_1 g_2 (g_1+g_2) + g^{3}_2 F_1}{(g_1+g_2)^3} \, ,
   \label{equation:fano_double_normal}
\eeqa
where $g_i = g_Q \sum_n \Gamma_{in}$ and
$F_i = (\sum_n \Gamma_{i n}(1-\Gamma_{i n})) /
\sum_n \Gamma_{i n}$, $i =$ 1 or 2.
From \refE{equation:fano_double_normal} the Fano factor for three
junctions in series can be easily obtained:
$F_{123}(g_1,F_1;g_2,F_2;g_3,F_2) =
F_{12}(g_{12},F_{12};g_3,F_3)$
with $g_{12} = g_1 g_1 /(g_1 +g_2)$.
This expression coincides with \refE{fano_incoh},
once we take into account the doubling of the charge.
Let us now consider the case when one of the two interfaces dominates
transport.
For $g^{\textrm{And}}_S \ll g_N$, N'-S junctions controls charge
transfer and it is thus not surprising to find that the Fano
factor is that of the N-S barrier alone:\cite{jong:1994}
$F_{\textrm{class}} = 2\, \sum_n R_{Sn}\,(1-R_{Sn})/{\sum_n R_{Sn}}$.
In the opposite limit of $g^{\textrm{And}}_S \gg g_N$, we have
instead the following result:
$F_{\textrm{class}} =\sum_n \Gamma_{Nn}\,(1-\Gamma_{Nn} /2) / {\sum_n \Gamma_{Nn} }$.
Note that it differs from the Fano factor for a single interface
of transparency distribution $\Gamma_{Nn}$.
Actually even if the resistance is dominated by the N'-N interface,
the presence of the N-S interface doubles the number of interfaces,
leading to this result.
Note also that for a completely transparent N'-N interface we
have a finite noise $F= 1/2$.
The conductance in this limit is $g_N$ [cf.
\refE{equation:cond_incoh}].
Again one could expect that $F$ should be zero, but actually
transport is slightly more subtle.
The effective system is that of a chaotic cavity connected through
two completely transparent interfaces of conductance $g_N$ to the
two leads.
The electron entering the cavity from the normal side has probability
1/2 of exiting from the same interface as an electron and 1/2 of
exiting as a hole on the other side.
In the second case the transferred charge is $2e$ with probability
1/2.
Thus the effective conductance is $g_N$, like for a normal Sharvin
contact, but with an effective Fano factor of 2 (for the charge) times
1/4 (for the $\Gamma(1-\Gamma)$ term).

\begin{figure}[tbh]
\centerline{
\includegraphics[scale=0.8]{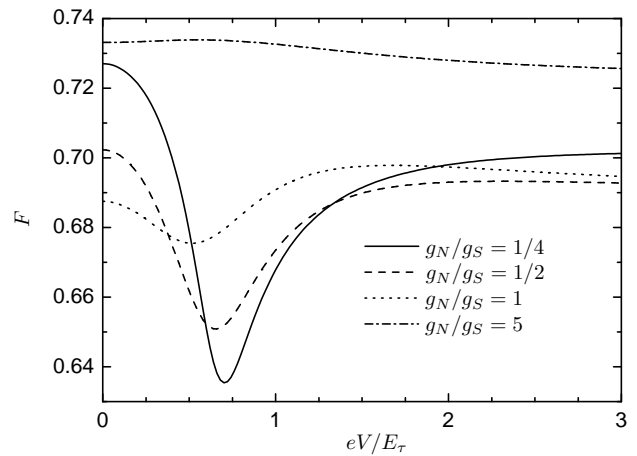}}
\caption{\label{fig:fano_dirty}
Differential Fano factor as a function of $eV / E_{\tau}$ for two
disordered interfaces at zero temperature. A dip appears for $eV
\approx E_{\tau}$ when $g_N \approx g_S$ and increases
for $g_N/g_S$ decrease.
}
\end{figure}

For intermediate values of the energy, noise, like the conductance,
has to be considered numerically.
Our results allow to study any situation.
We plot in Fig. \ref{fig:fano_dirty} the Fano factors
for the same parameters considered for the conductance
in Fig. \ref{fig:cond_dirty}.
The qualitative behavior resembles that of the noise in
long diffusive structures.
In particular a minimum at finite voltage for the
Fano factor is present when $g_S \gg g_N$.
This is very similar to the minimum in the differential
Fano factor for a wire in good contact with normal
and superconducting reservoirs.\cite{belzig:2002,kozhevnikov:2000}

\subsection{Effect of the channel distribution on current and noise}
\label{barriers}

\begin{figure}
\centerline{
\includegraphics[scale=0.8]{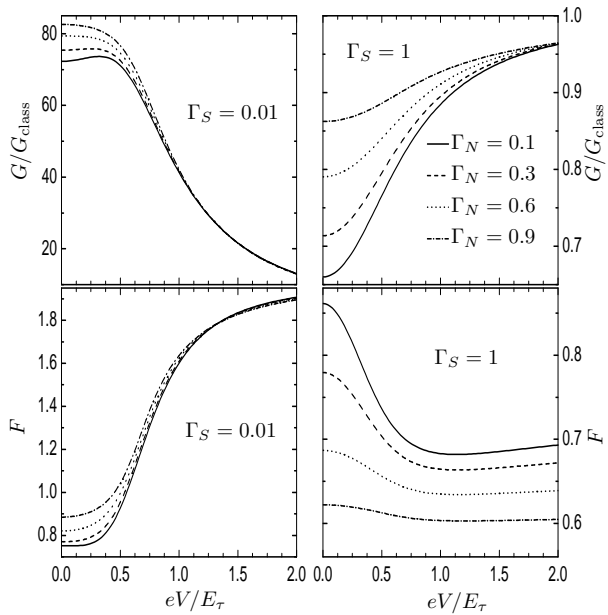}}
\caption{\label{Chan1}
Differential conductance and Fano factor as a function of $
eV/E_{\tau} $ at zero temperature. 
Channel transparencies of the two interfaces have unimodal
distribution. 
In the left panel, N'-S junction is tunnel ($\Gamma_S= 0.01$) and in
the right panel N'-S junction is transparent ($\Gamma_S = 1$).  }
\end{figure}
\begin{figure}[t]
\centerline{
\includegraphics[scale=0.8]{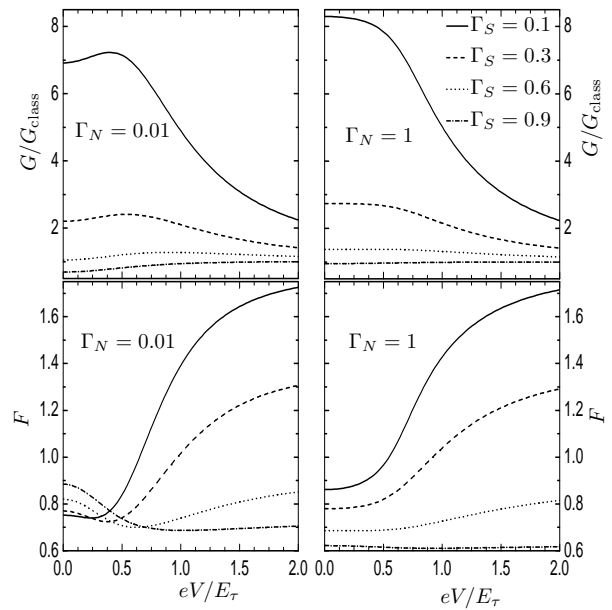}}
\caption{\label{Chan2}
Differential conductance and Fano factor as a function of $eV/E_{\tau}
$ at zero temperature. 
Channel transparencies of the two interfaces have unimodal
distribution. 
In the left panel, N-N' junction is tunnel ($\Gamma_N= 0.01$) and in
the right panel N-N' junction is transparent ($\Gamma_N = 1$).  }
\end{figure}

Let us now consider the genuine effect of the
modification of the channel distribution.
The optimal situation is when the normal-state conductances of the two
interfaces are equal, so that the dependence on the distribution 
of the channel transparencies of both interface should be maximum.
For simplicity in the presentation we will discuss only the case of
$\Gamma_n=\Gamma$ for all $n$.
We thus vary the transparency and the number of channels at each interface 
in such a way that the ratio of the
normal-state conductance is kept equal to 1.
Note when the channels are transparent this does not 
mean necessary that the contact region must be very small (to keep
the same conductance of the tunneling case). 
It is enough that the distribution of channels of a large 
junction is bimodal, with a large majority of 
the channels completely opaque ($\Gamma=0$) 
and few of them of the given transparency.

We calculated the energy dependence of the
conductance and of the noise as a function of the
transparency of the interfaces.
Results are reported in \refF{Chan1} and \refF{Chan2}.
In \refF{Chan1} we set $\Gamma_S=0.01$ (left pannels) and
$\Gamma_S=1$ (right pannels) and we vary $\Gamma_N$
from 0.1 to 0.9.
In \refF{Chan1} we plot the same curves exchanging the role
of $\Gamma_N$ and $\Gamma_S$.

First we note how important is the channel transparency to
predict the value of both the conductance and the noise.
Knowledge of the conductance alone is not enough.
Once the conductance is known, the energy dependence of both current
and noise can give valuable indications on the channel distribution.
A second important remark is the qualitatively similar
behavior of the conductance and the Fano factor.
This is particularly evident for the case when $\Gamma_N\gg\Gamma_S$
(left panel of \refF{Chan1}).
The differential Fano factor is linearly related to
the conductance, 
$F(E) \approx \gamma_0 -\gamma_1 G(E)$, 
with $\gamma_0$ and $\gamma_1$ positive constants.
This behavior was proved analytically for tunneling contacts
between a superconductor and a wire in Ref. \onlinecite{houzet:2004}.
Actually this behavior seems to be a general property  of the 
whole set of plots, with variable accuracy.
The differential Fano factor  looks like the differential conductance
upside down.
This is only a qualitative behavior, the proportionality factor
depends on the actual transparency, as was found in
Ref. \onlinecite{houzet:2004}.

\section{Conclusions}
\label{conclusions}

We studied the energy dependence of the current noise in a double barrier
N-N'-S structure for arbitrary transparency of the barriers. 
In particular, we could describe the crossover between the completely
coherent and incoherent regimes. 
The noise in the double-barrier structure ressembles the one found
earlier in an extended geometry, like a wire. 
Namely, we found that the energy dependence of the current and noise are
qualitatively strongly related though quantitatively independent.
The distribution of transparencies at the barriers strongly
influences both the current and noise.
We suggest that a measurement of both quantities in the same sample
would provide valuable information on the properties of the
interfaces.

\acknowledgements

We thank F. Lefloch for stimulating discussions. 
We acknowledge financial support from IPMC of Grenoble (F.P.) 
and from the French ministry of research through contract 
ACI-JC no 2036 (M.H.).

\appendix

\newcommand{\chu}{{\tilde c}_1}
\newcommand{\chd}{{\tilde c}_2}
\newcommand{\cu}{c_1}
\newcommand{\cd}{c_2}

\newcommand{\shu}{{\tilde s}_1}
\newcommand{\shd}{{\tilde s}_2}
\newcommand{\su}{s_1}
\newcommand{\sd}{s_2}

\section{Fano factor equations}

In this appendix we give the explicit expressions for ${\cal S}$ and
the coefficients $a$, $b$, and $d$, entering the definition of $\check
\phi$. We use the shorthand notation $\cosh \theta_1 = \chu$ , 
$\sinh \theta_1 = \shu$, $\cos \theta_2= \cd$, and $\sin \theta_2 = \sd$.

We begin with the three coefficients:
\begin{widetext}
\beq
a = \frac{g_N \, \qav{A_0} + \chu \left( -2 \, g_N \, c \,
\qav{A_1} + c^2 ( g_N \, \qav{A_{1}} + g_S \, \qav{A_{2}} + 2 \, i \,
g_D \, \varepsilon \, \sd ) \right) }{\qav{B}}
\eeq
with
\begin{subequations}
\beqa
A_0 & = &  i \sinh \theta \, \sd^{2} \, 
\overline{Z_N}\,{Z_N}^{2}{\Gamma_N}^{2}
 + \Big( \big( 
2\, i \sinh \theta  \, \chu \cd  - i \chu \shu + \cd \sd  \big) {Z_N}^{2}
 -2\,\sd \cd \overline{Z_N} Z_N \Big) \Gamma_N
- 2  i \sinh \theta \,  Z_N \nonumber \, , \\
A_1 & = &
i \, \sinh \theta \, \chu \sd
\overline{Z_N}\,{Z_N}^{2}{\Gamma_N}^{2}
 + \left( i \, \sinh \theta \, \cd {Z_N}^{2}
-2\,\chu \cd \sd \overline{Z_N} \, Z_N
\right) \Gamma_N +2\,Z_N\,\sd \nonumber \, ,  \\
A_{2} & = & 
- \cosh \theta \, \cd^2 \, \chu 
 \overline{Z_S}\,{Z_S}^{2}{\Gamma_S}^{2} 
 + \left( \cosh \theta \, \chd \,
{Z_S}^{2} +2\,\chu \sd \cd \overline{Z_S}\,Z_S \right) \Gamma_S  + 2\,Z_S\,\cd
\nonumber \, ,  \\
B & = & i g_S \,{Z_S}{\Gamma_S}^{2} \cosh^2 \theta -i g_N
{Z_N}^{2}\,{\Gamma_N} \sinh^2 \theta -2\, g_S \,\,Z_S\, \sinh
\theta -2\,g_D\,\varepsilon\, \cosh \theta +2\,i g_N \,Z_N\, \cosh
\theta
\, .  \nonumber 
\eeqa
\end{subequations}

Then
\beqa
b & = & c\, (1 - 2 f^{2}_{L0} ) + f^{2}_{T0} b_T
\eeqa
with

\beq
b_T = -2 c^3 + \frac{g_N}{ 4 \varepsilon g_D} \left[ \qav{\beta_0} c^2
\frac{\shu }{ \sd} -2 c\, \left( \qav{\beta_0} \frac{ \shu }{ \sd } + \frac{a_2 }{ \tan \theta_2} \right) + \qav{\beta_1}
\frac{a_1 }{ \chu \sd} + \qav{\beta_2} a_2 \frac{\shu }{ \chu } + \qav{\beta_3} \frac{\shu }{ \sd} \right]\,,
\eeq
and
\begin{subequations}
\beqa
\beta_0 & = &
- \left( \chu - \cd \right)^2  | Z_N |^2\,{\Gamma_N}^{3}
- 4 \left( 1+\chu \cd -2\, \cd^{2} \right) | Z_N |^2\,{\Gamma_N}^{2}+4 \left( 1-2\,
\cd^{2} \right) | Z_N |^2\,\Gamma_N
\nonumber  \\
\beta_1
& = &  \left( \chu -\cd \right)^2 \left( 1 - \chu \, \cd \right) | Z_N |^2\,{\Gamma_N}^{3}
 + 2\left( \cd   \chu^{3}-4\,
 \cd ^{2}-5\, \chu^{2}+2+4\,\chu \cd +1\,
 \cd ^{3}\chu \right) | Z_N |^2\,{\Gamma_N}^{2}
\nonumber \\
&& +4 \left( 2\, \cd^{2}-3\,\chu
\cd +2\,  \chu^{2}-1 \right) | Z_N
|^2\,\Gamma_N+8\,| Z_N |^2\,\chu \cd \nonumber \\ 
\beta_2 &
=& - \left( \chu - \cd \right)^2 | Z_N
|^2\,{\Gamma_N}^{3} - 2 \left( 2-\, \cd^{2}
-\, \chu^{2} \right) | Z_N
|^2\,{\Gamma_N}^{2}  +12\,| Z_N |^2\,\Gamma_N-8\,| Z_N|^2
\nonumber \\
\beta_3 & = &
- \left( \chu - \cd \right)^2 | Z_N|^2
\, {\Gamma_N}^{3} -4 \left( 1 + 4\,\chu \cd -2\,
 \cd^{2} \right) | Z_N |^2\,{\Gamma_N}^{2}
 +4 \left( 1-2\, \cd^{2} \right) | Z_N
|^2\,\Gamma_N
\, .
\nonumber
\eeqa
\end{subequations}
Finally
\beqa
d & = & - 2 f_{L0} f_{T0} \left( 1 + c^2 + a_2 \tan \theta_2 \right)
\,.
\eeqa

The explicit form of $\mathcal{S}_T (E)$ reads:
\beq
\mathcal{S}_T (E)
=
\mathcal{G} ( E) \,  \frac{ b_T}{ 2c}  + \frac{c \, g_S \,
\chu}{8}
\left[
 \left( \qav{\alpha_1} \, a_1 \, \shu + \qav{\alpha_2} \, a_2 \, \cos
 \theta_2 \right) +\qav{\alpha_3}\,  c^2 
\right]
\eeq
with
\begin{subequations}
\beqa
\alpha_1 & =&
- \left( \chu + \sd \right)^2 |Z_S|^2\,{\Gamma_S}^{3}
+4 \left( \chu \sd -2\, \cd^{2} + 1\right) |Z_S|^2 \, {\Gamma_S}^{2}
+4  \left( -1+2\,  \cd^{2} \right) |Z_S|^2\,\Gamma_S
\nonumber
\, , \\
\alpha_2 & = &
- \left( \chu + \sd \right)^2
 |Z_S|^2\,{\Gamma_S}^{3}+
2 \left( -1-\, \cd^{2}+\,
\chu^{2} \right) |Z_S|^2\,{\Gamma_S}^{2}
 +12\,\Gamma_S\,|Z_S|^2-8\,|Z_S|^2
\nonumber \, ,  \\
\alpha_3
& = &
\sd \left( \chu + \sd \right)^2 |Z_S|^2\,{\Gamma_S}^{3}
 + 2 \left(  \cd^{2}\sd
-4\,\chu - \chu^{2}\sd -3\,\sd \right) |Z_S|^2\,{\Gamma_S}^{2}
+ 4 \left( 3\,\sd +2\,\chu \right)
|Z_S|^2\,\Gamma_S-8\,|Z_S|^2\,\sd
\, .
 \nonumber
\eeqa
\end{subequations}

\end{widetext}

\end{document}